\documentclass{elsart}

\usepackage{amsmath}    % eg. \text
\usepackage{amssymb}    % eg. \Box, \gtrapprox, \gtrsim, \less*

\usepackage{graphicx}

\begin{document}

\begin{frontmatter}

\title{Modeling viscoelastic flow with discrete methods}

\author{Ell\'ak Somfai},
\ead{ellak@lorentz.leidenuniv.nl}
\author{Alexander N.\ Morozov},
\author{Wim van Saarloos}
\address{Instituut--Lorentz, Universiteit Leiden, PO Box 9506, 2300 RA Leiden,
    Netherlands}

\begin{abstract}
The hydrodynamics of viscoelastic materials (for example polymer melts and
solutions) presents interesting and complex phenomena, for example
instabilities and turbulent flow at very low Reynolds numbers due to normal
stress effects and the existence of a finite stress relaxation time.  This
present work is motivated by renewed interest in instabilities in polymer
flow.  The majority of currently used numerical methods discretize a
constitutive equation on a grid with finite difference or similar methods.  We
present work in progress in which we simulate viscoelastic flow with
dissipative particle dynamics.  The advantage of this approach is that many of
the numerical instabilities of conventional methods can be avoided, and that
the model gives clear physical insight into the origins of many viscoelastic
flow instabilities.
\end{abstract}

\end{frontmatter}

%%%%%%%%%%%%%%%%%%%%%%%%%%%%%%%%%%%%%%%%%%%%%%%%%%%%%%%%%%%%%%%%%%%%%%%%%%%%%%%

\section{Introduction}
\label{sec:introduction}

\begin{figure}
\centering\includegraphics[width=0.4\textwidth,clip]{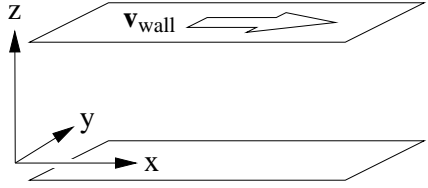}
\caption{\label{fig:couettecoord}
    The plane Couette flow geometry of our simulations.}
\end{figure}

One of the most important properties of viscoelastic fluids is that shear flow
affects not only the off-diagonal (shear) component of the stress tensor, but
also the diagonal elements: they change with respect to each other
\cite{larson88,bird87v1}.  In the plane Couette flow geometry of
Fig.~\ref{fig:couettecoord}, where the viscosity $\eta$ is 
\begin{equation}
\sigma_{xz} = \eta \dot\gamma 
\end{equation}
($\sigma_{\alpha\beta}$ is the stress tensor and $\dot\gamma=\partial
v_x / \partial z$ is the shear rate), the first normal stress difference
$N_1$ is defined as 
\begin{equation}
\sigma_{xx}-\sigma_{zz} = N_1(\dot\gamma) = \Psi_1 \dot\gamma ^2 \,,
\end{equation}
where the first normal stress coefficient $\Psi_1$ is finite for small
shear rate $\dot\gamma$.

Many of the unusual flow phenomena of viscoelastic fluids can be traced back
to this nonzero normal stress difference.  The extra forces generated by the
viscoelastic stresses often destabilize the flow, leading to instabilities.
This work has been motivated by the realization that many key issues regarding
viscoelastic instabilities are unresolved. Eg., it was suggested recently that
while the viscoelastic Poiseuille flow might be linearly stable, it could be
nonlinearly unstable, and this might be a route to melt-fracture type behavior
\cite{meulenbroek03,meulenbroek04}.
These viscoelastic instabilities occur even at vanishing Reynolds numbers, and
are driven not by kinetic forces but by elastic forces.  The control parameter
is the Weissenberg number \textsf{Wi}, the ratio of the characteristic
relaxation time of the fluid ($\tau$) and the characteristic time of the flow:
for a shear flow it is $\dot\gamma\tau$.  As \textsf{Wi} is increased, the
instabilities lead to a complex non-stationary flow, often called viscoelastic
turbulence \cite{larson00,groisman00}, based on similarities to the Reynolds
number driven turbulence. 

Clearly numerical methods are in great need to understand these complex
phenomena.  However, the classical rheological engineering approach, based on
finite volume or finite difference discretization of the Navier-Stokes
equation together with the viscoelastic constitutive equations, runs into
numerical instabilities in what is commonly referred to as ``high Weissenberg
number problem'' \cite{owens02}.  Our goal is to overcome this barrier by
turning to alternative, discrete methods of fluid dynamics.

In this paper we describe an extension of one of the successful discrete
methods, the dissipative particle dynamics, to viscoelastic fluids.  In this
approach we don't start from a closed set of equations, instead use kinetic
considerations. This is work in progress, and we show the first validation
results.

%%%%%%%%%%%%%%%%%%%%%%%%%%%%%%%%%%%%%%%%%%%%%%%%%%%%%%%%%%%%%%%%%%%%%%%%%%%%%%%

\section{Dissipative particle dynamics}

In the method of dissipative particle dynamics (DPD)
\cite{hoogerbrugge92,espanol95} the fluid is represented with particles, each
one corresponding to a macroscopic blob of the fluid.  The particles interact
with each other via finite range pairwise forces.  The force exerted on
particle $i$ of a Newtonian fluid can be written as a sum of conservative,
dissipative and random contributions:
\begin{equation}
\label{eq:force}
\mathbf{f}_i = \sum_{j\not=i} \left(
\mathbf{f}_{ij}^\text{cons} + \mathbf{f}_{ij}^\text{diss}
  + \mathbf{f}_{ij}^\text{rand}
\right) \,.
\end{equation}

The conservative part of the force is a soft repulsion:
\begin{equation}
\mathbf{f}_{ij}^\text{cons} = \left\{ 
\begin{array}{ll}
    a \left(1-r_{ij}/r_c\right)\hat{\mathbf{r}}_{ij}, \quad & r_{ij} < r_c \\
    0, & r_{ij}\ge r_c
\end{array}
\right.
\end{equation}
where $\mathbf{r}_{ij}=\mathbf{r}_i - \mathbf{r}_j$ is the separation vector
of the particles, with distance $r_{ij}=|\mathbf{r}_{ij}|$, and unit vector
$\hat{\mathbf{r}}_{ij}=\mathbf{r}_{ij}/r_{ij}$.

The dissipative part of the force acts to equalize velocities of nearby
particles. It is also central force:
\begin{equation}
{\mathbf f}_{ij}^\text{diss} = -\gamma\, w^\text{diss}(r_{ij})\,
(\hat{\mathbf{r}}_{ij}\cdot\mathbf{v}_{ij})\,\hat{\mathbf{r}}_{ij} \,.
\end{equation}

The random part of the force represents a coupling to a heat bath:
\begin{equation}
\mathbf{f}_{ij}^\text{rand} = \sigma\, w^\text{rand}(r_{ij})\,
\xi_{ij}\,\hat{\mathbf{r}}_{ij} \,,
\end{equation}
where $\xi_{ij}=\xi_{ji}$ is a Gaussian random variable, independent for each
$ij$ pair of particles and timestep, with zero mean and $\Delta t^{-1}$
variance.

The coefficients $\gamma$ and $\sigma$ and the two weight functions cannot be
chosen arbitrarily: in order that the fluctuation-dissipation theorem holds
\cite{espanol95}, they must be related by
\begin{equation}
w^\text{diss}(r) = \left[w^\text{rand}(r)\right]^2 \,,
\end{equation}
\begin{equation}
\sigma^2 = 2\gamma k_\text{B}T \,.
\end{equation}

We use the following weight function \cite{chen04}:
\begin{equation}
w^\text{diss}(r) = \left[w^\text{rand}(r)\right]^2 = \left\{
\begin{array}{ll}
    \sqrt{1 - r/r_c},\quad & r<r_c \\
    0, & r>r_c
\end{array}
\right.
\end{equation}

Given the interparticle forces, Newton's equations of motion are solved with a
version of the velocity-Verlet algorithm \cite{groot97}.
As customary, our units were the cutoff length $r_c$ and the mass of a
particle.

In our plane Couette flow geometry (see Fig.~\ref{fig:couettecoord}) we have
periodic boundary conditions in the $x$ (streamwise) and $y$ (spanwise)
direction, and no-slip walls perpendicular to the $z$ (gradient) axis.  The
walls are implemented as a soft repulsion potential in the normal direction:
\begin{equation}
\mathbf{f}^\text{wall}_z =  \left\{
\begin{array}{lll}
    -a^\text{wall} z,       & z < 0             &\\
    0,                      & 0\le z\le L\quad  & \text{(in bulk)} \\
    a^\text{wall}(z-L),\quad&  z>L              &
\end{array}
\right. \,,
\end{equation}
where $L$ is the distance between the walls.  To realize no-slip boundary
conditions, at each timestep we update the particles' velocity component
parallel to the walls as
\begin{equation}
\mathbf{v}_\| \; \Leftarrow \; \mathbf{v}_\| +
    \alpha(z)(\mathbf{v}^\text{wall} - \mathbf{v}_\|)  \,,
\end{equation}
where the factor $\alpha(z)$ is selected so that the velocity is unaffected in
the bulk, the wall velocity is imposed upon particles well outside the walls,
and there is a continuous crossover between these limits.  Near the bottom
wall it is 
\begin{equation}
\alpha(z) = \left\{
\begin{array}{lll}
    0,              & z \ge 0         &\text{(in bulk)} \\
    -z/d_0,\quad    & -d_0<z<0\quad &\text{(near wall)} \\
    1,              & z \le -d_0      &\text{(well outside wall)}
\end{array}
\right. 
\end{equation}
and similarly defined near the top wall.  This approach is admittedly not
elegant, as for example it has a non-trivial timestep dependence, but
nevertheless achieves no-slip boundary conditions with minimal effort.

With these boundary conditions the particles wander outside the nominal wall
position to a limited distance depending on the temperature and pressure.  For
too sharp walls the density near the walls displays oscillations as a function
of distance from the wall.  These are probably the consequence of local
crystallization near a sharp surface.  This unphysical phenomenon is avoided
by softening the wall; we used $a^\text{wall} = a$.

The $z$ components of the stress tensor at the walls are readily available
from the momentum transfer between the wall and the particles.  In the bulk,
however, they have to be computed from the pairwise forces
\cite{goldhirsch02}:
\begin{equation}
\label{eq:gold}
\sigma_{\alpha\beta} = \frac{1}{V} \left\{
-\sum_{\langle i,j\rangle} \mathbf{f}_{ij,\alpha} \mathbf{r}_{ij,\beta}
-\sum_i m_i(\mathbf{v}_i-\bar{\mathbf{v}})_\alpha (\mathbf{v}_i
-\bar{\mathbf{v}})_\beta
\right\} \,.
\end{equation}

We validated our code for Newtonian fluids by calculating the stresses with
different methods: at the walls, the global bulk value
(where $\bar{\mathbf{v}}$ is taken as the average velocity of particles at the
same height $z$), and calculated locally [here Eq.~(\ref{eq:gold}) is
integrated with a coarse graining kernel, which vanishes outside the 
neighborhood of a point].  All measurements were equal within error except the
local stress, which differed by 10\%---we attribute this to the difference in
the contribution of fluctuations.
We then calculated the viscosity from the shear stress, which compared well
with an independent measurement from the relaxation time of the shear velocity
profile's build-up at sudden start-up of the shear boundary conditions from
rest.

%%%%%%%%%%%%%%%%%%%%%%%%%%%%%%%%%%%%%%%%%%%%%%%%%%%%%%%%%%%%%%%%%%%%%%%%%%%%%%%

\section{DPD for viscoelastic fluids}

To simulate viscoelastic hydrodynamics we connected DPD particles with
springs: some fraction (in this paper all) of the particles are paired up to
form dumbbells, so Eq.~(\ref{eq:force}) receives an extra term ${\mathbf
f}^\text{dumb}$.  Linear (Hookean) springs give rise to nonphysical effects in
elongational flow \cite{bird87v1}, therefore we used finite extensible
non-linear elastic (FENE) springs:
\begin{equation}
\mathbf{f}^\text{dumb}(\mathbf{r}) = \frac{H \mathbf{r}}
{1 - (r/r_\text{max})^2}
\end{equation}
where $\mathbf{r}$ is the separation between the endpoints of the dumbbell,
$H$ is the Hookean spring stiffness (at zero extension), and $r_\text{max}$ is
the maximum extension of the spring.

\begin{figure}
\centering\includegraphics[height=0.31\textwidth,clip]{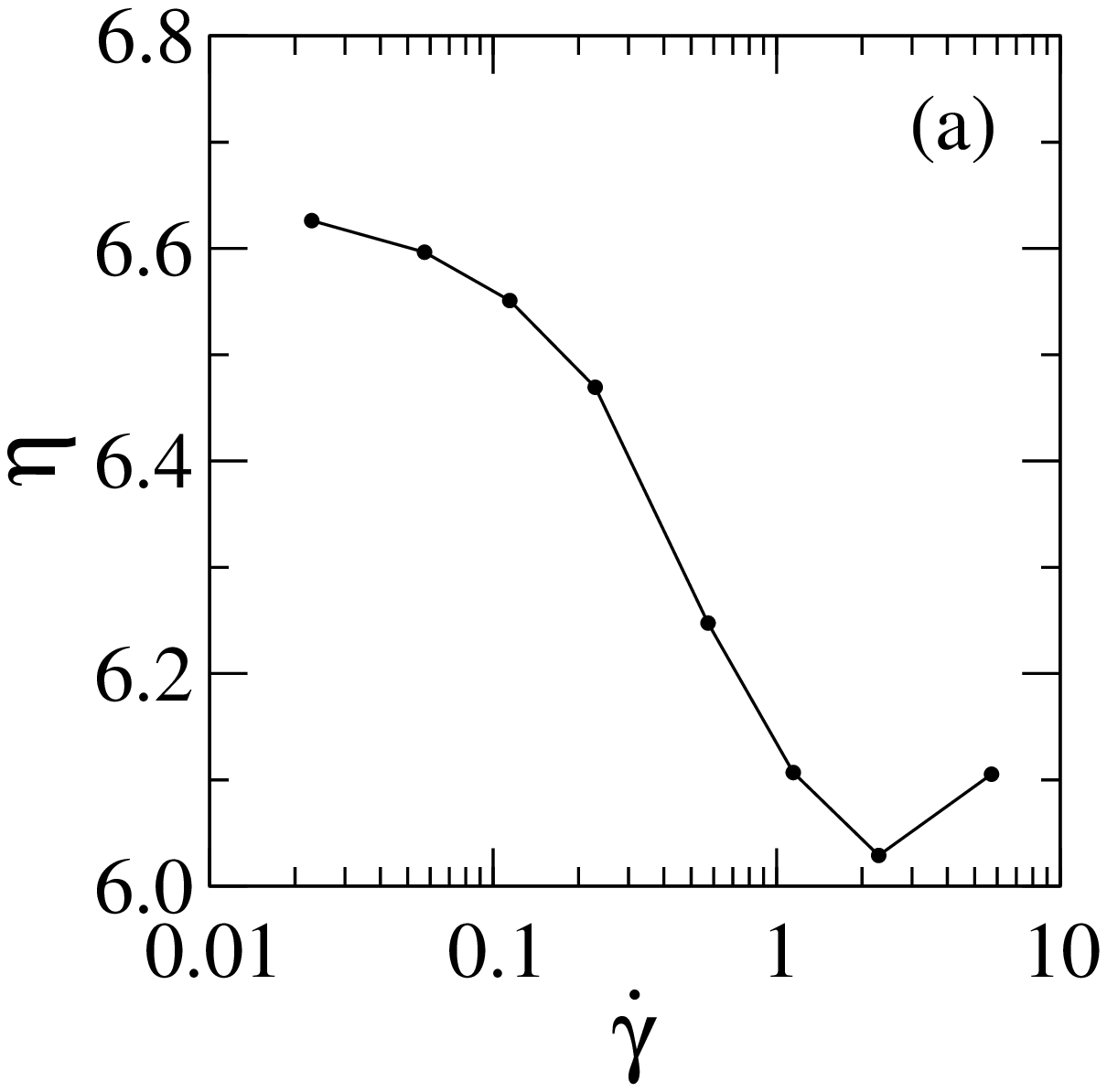}%
\hfill%
\centering\includegraphics[height=0.31\textwidth,clip]{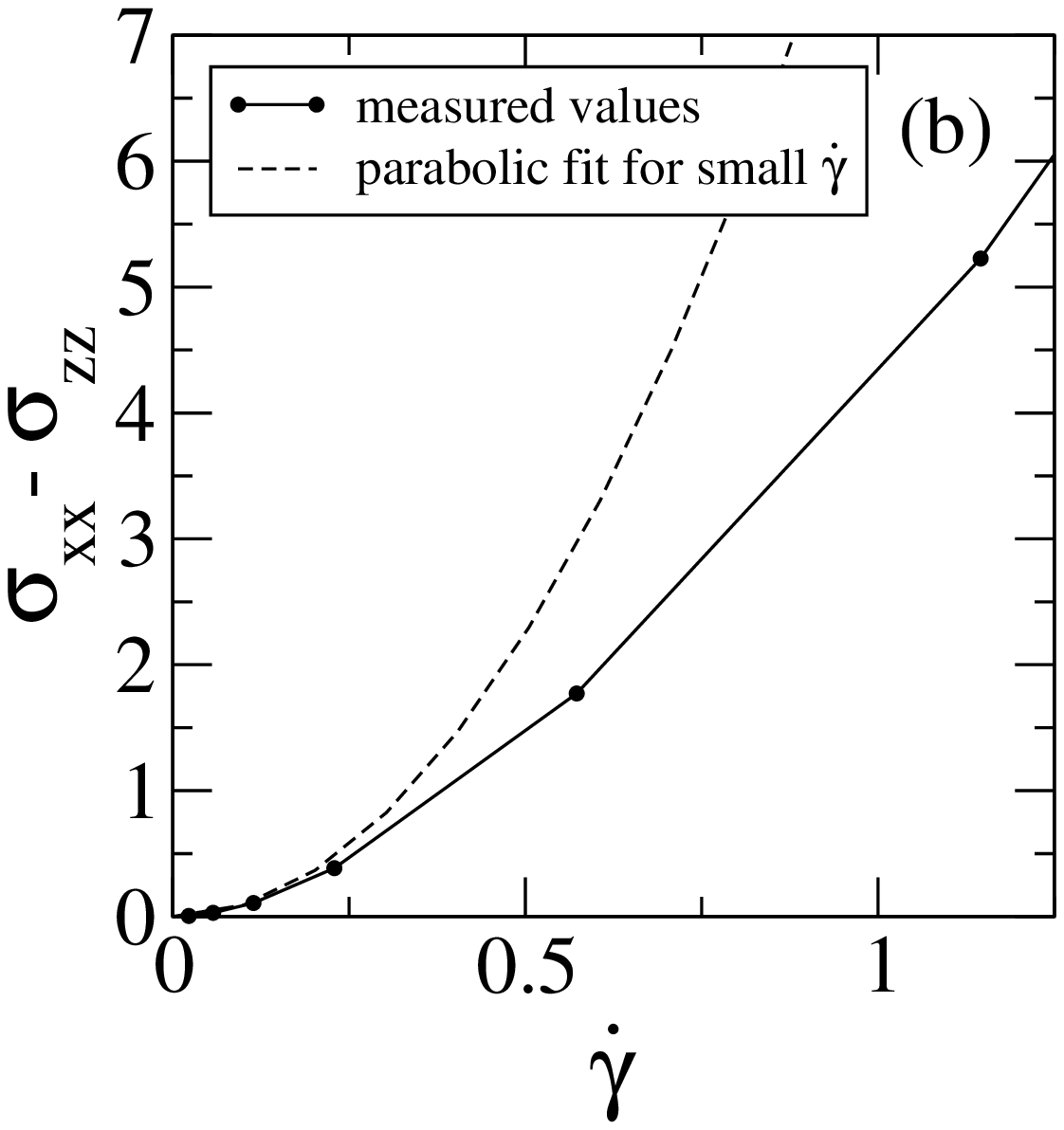}%
\hfill%
\centering\includegraphics[height=0.31\textwidth,clip]{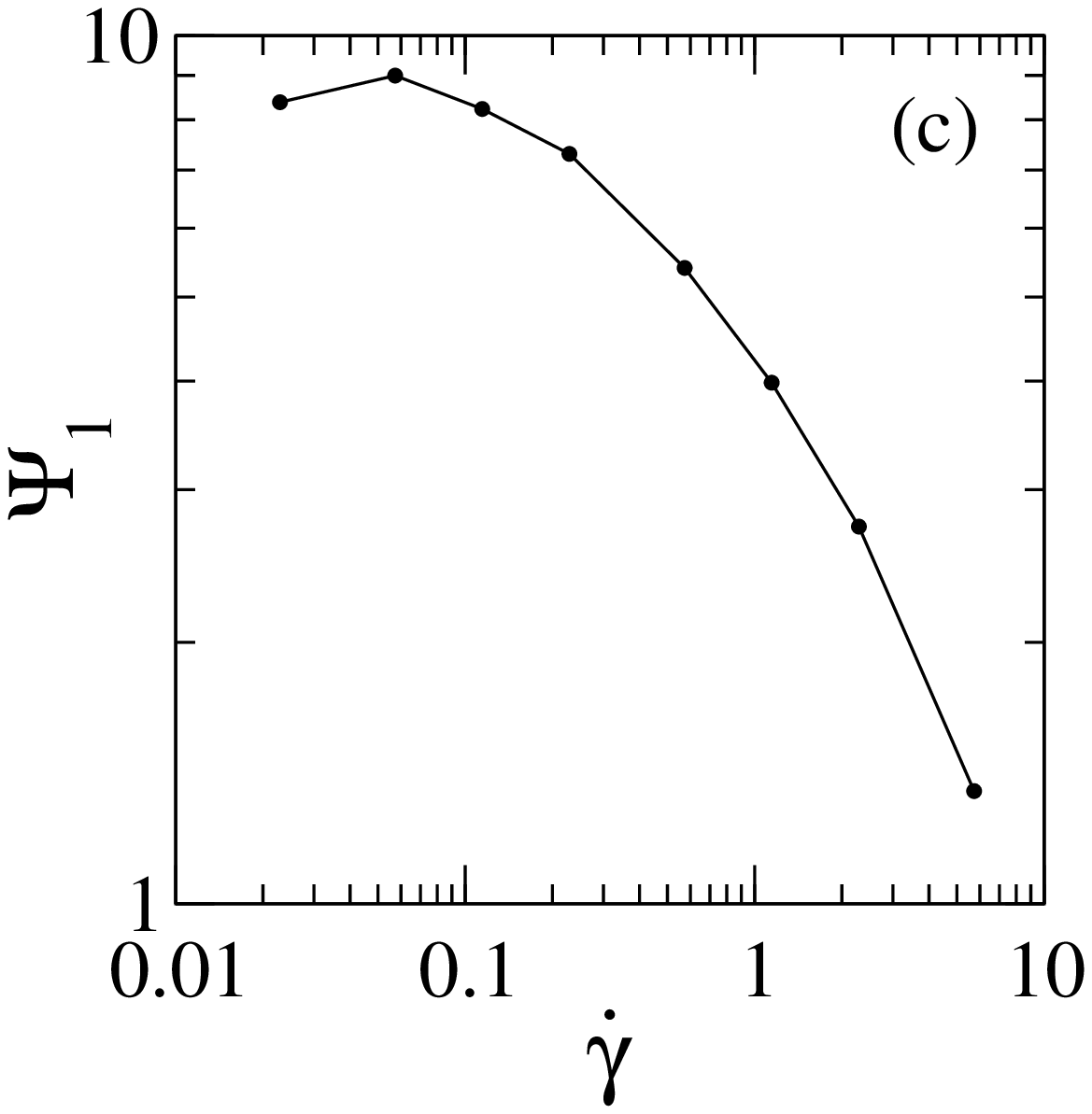}
\caption{\label{fig:etapsi}
    \textbf{(a)} Viscosity as a function of the shear rate. For large shear
rates the viscosity decreases slightly (shear thinning).
\textbf{(b)} First normal stress difference and \textbf{(c)} first normal
stress coefficient.  The normal stress difference is quadratic for small shear
rate $\dot\gamma$, but the coefficient decreases significantly for
$\dot\gamma\gtrsim 1$.}
\end{figure}

As the first validation test, we measured the velocity profile in our shear
cell, and obtained the expected linear profile.  The shear viscosity, however,
is now weakly dependent on the shear rate, as shown on
Fig.~\ref{fig:etapsi}a.

We also measured the first normal stress difference, see
Fig.~\ref{fig:etapsi}b.  The figure shows that the model does achieve its
goal, namely that it yields a normal stress difference, which for small
$\dot\gamma$ increases as $\dot\gamma^2$.  At shear rate $\dot\gamma\gtrsim
1$, the coefficient $\Psi_1$ starts to decrease notably.

\begin{figure}
\centering\includegraphics[width=0.8\textwidth,clip]{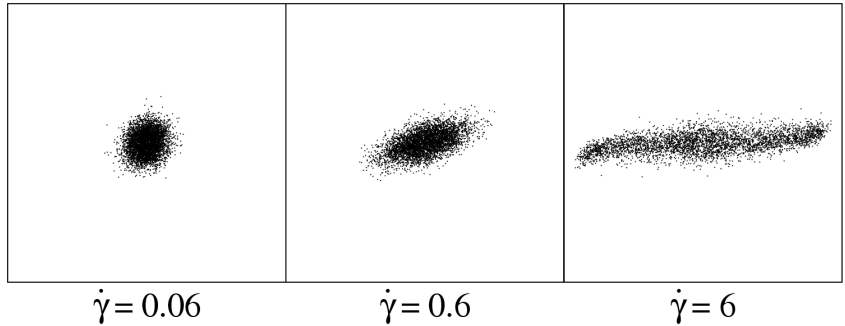}
\caption{\label{fig:dumbbells}
    Distribution of dumbbell configurations in the shear plane.  Each dot
corresponds to the end-to-end vector of a dumbbell in the $xz$ plane.  For
small $\dot\gamma$ the orientation is isotropic, at moderate shear rates it is
an elongated ellipse, and at large shear rates it is further deformed.  The
sides of the box represent the maximum elongation $r_\text{max}$ of the FENE
springs.}
\end{figure}

The viscoelastic modes are represented by the FENE dumbbells, to which we have
full access in the numerical simulations.  Figure~\ref{fig:dumbbells} shows
the distribution of dumbbell orientation and extension.  At low shear rates
the dumbbells are isotropically oriented.  At larger shear rates the dumbbells
become elongated, with the typical orientation at a small angle with the
stream direction.  At the highest shear rate the distribution shows almost
zero angle of the dumbbells with respect to the stream lines.

\begin{figure}
\centering\includegraphics[width=0.8\textwidth,clip]{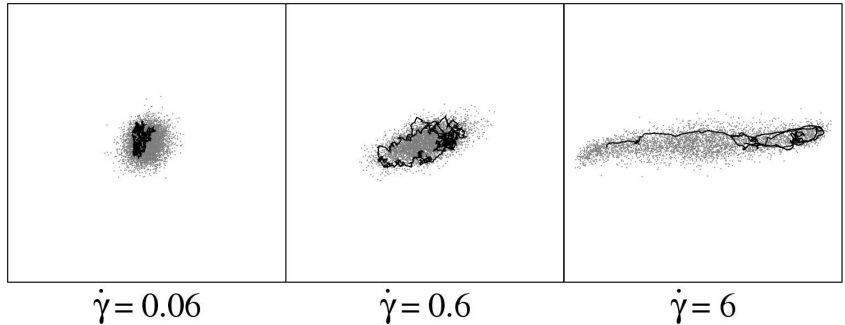}
\caption{\label{fig:traj}
    Trajectories of the configuration of a dumbbell at different shear rates.
A selected dumbbell is traced for 50 time units for the two smaller shear
rates, and 20 time units for the largest shear rate.}
\end{figure}

The time evolution of a typical dumbbell orientation is plotted on
Fig.~\ref{fig:traj}.  For small shear rates the trajectory resembles a random
walk in the configuration space, while at larger shear rates the dumbbells
tumble: if by fluctuation the two endpoints get to different height $z$, they
are dragged with different velocities, resulting in a horizontal stretch.  If
at this point the endpoints move to the same height, the differential drag
disappears, and the dumbbells can relax.  This is augmented with the
rotational component of the shear flow to yield a tumbling motion.

%%%%%%%%%%%%%%%%%%%%%%%%%%%%%%%%%%%%%%%%%%%%%%%%%%%%%%%%%%%%%%%%%%%%%%%%%%%%%%%

\section{Summary}

In conclusion, we presented preliminary results on extending DPD for
viscoelastic fluids by connecting DPD particles to form dumbbells.  The model
shows (shear rate dependent) normal stress difference, and a small amount of
shear thinning.  This is our first step to study numerically the instabilities
and turbulence in viscoelastic fluids with a method complementary to direct
numerical simulations of the constitutive equations.

%%%%%%%%%%%%%%%%%%%%%%%%%%%%%%%%%%%%%%%%%%%%%%%%%%%%%%%%%%%%%%%%%%%%%%%%%%%%%%%

% The Appendices part is started with the command \appendix;
% appendix sections are then done as normal sections
% \appendix

% \section{}
% \label{}

\begin{ack}
This research has been supported by the PHYNECS training network of the
European Commission under contract HPRN-CT-2002-00312.
We thank Mar Serrano for useful discussions.
\end{ack}

\bibliographystyle{elsart-num}
\bibliography{fluidref}

\end{document}